\documentclass[12pt,preprint]{aastex}

\usepackage{emulateapj5}

\begin{document}

\title{The effects of thermal conduction on the hot accretion flows}

\author{Mohsen Shadmehri and Fazeleh Khajenabi
\affil{Department of Physics, School of Science, Ferdowsi
University, Mashhad,
Iran}}\email{mshadmehri@science1.um.ac.ir,fkhajenabi@science1.um.ac.ir}

\begin{abstract}
Thermal conduction has been suggested as a possible mechanism by
which sufficient extra heating is provided in radiation-dominated
accretion flows. We consider the extreme case in which the
generated energy due to the viscosity  and the energy transported
by a saturated form of thermal conduction are balanced by the
advection cooling. For the steady-state structure of such
accretion flows a set of self-similar solution are presented.
Based on these solutions while the radial and the rotational
velocities are both sub-Keplerian, increasing the level of
thermal conduction has the effects of decreasing the rotational
velocity, but increasing the radial velocity. Conduction provides
extra heating and the temperature of the gas increases with
thermal conduction.

\end{abstract}


\section{Introduction}

The thin accretion disk model describes flows in which the
viscous heating of the gas radiates out of the system immediately
after generation (Shakura $\&$ Sunyaev 1973). However, another
kind of accretion has been studied during recent years where
radiative energy losses are small so that most of the energy is
advected with the gas. These Adection-Dominated Accretion Flows
(ADAF) occur in two regimes depending on the mass accretion rate
and the optical  depth. At very high mass accretion rate, the
optical depth becomes very high and the radiation can be trapped
in the gas. This type of accretion which is known under the name
'slim accretion disk' has been studied in detail by Abramowicz et
al. (1988). But when the accretion rate is very small and the
optical depth is very low, we may have another type of accretion
(Narayan \& Yi 1994; Abramowitz et al. 1995; Chen 1995). However,
numerical simulations of radiatively inefficient accretion flows
revealed that low viscosity flows are convectively unstable and
convection strongly influences the global properties of the flow
(e.g., Igumenshchev, Abramowicz, $\&$ Narayan 2000). Thus, another
type of accretion flows has been proposed, in which convection
plays as a dominant mechanism of transporting angular momentum
and the local released viscous energy (e.g., Narayan,
Igumenshchev, $\&$ Abramowicz 2000).

This diversity of models  tells us that modeling the hot
accretion flows is a challenging and controversial problem. We
think, one of the largely neglected physical ingredient in this
field, is {\it thermal conduction}. But a few authors tried to
study the role of "turbulent" heat transport in ADAF-like flows
(Honma 1996; Manmoto et al. 2000) Since thermal conduction acts to
oppose the formation of the temperature gradient that causes it,
one might expect that the temperature and density profiles for
accretion flows in which thermal conduction plays a significant
role to appear different compared to those flows in which thermal
conduction is less effective. Just recently, Menou (2005) studied
the effect of saturated thermal conduction on optically thin
ADAFs using an extension of self-similar solutions of Narayan
$\&$ Yi (2004). His solutions suggest that thermal conduction may
be an important physical factor to understand hot accretion onto
dim accreting black holes.

The aim of the work presented here is to investigate whether
thermal conduction can affect  the dynamics of an optically {\it
thick} radiation-dominated accretion flow. In \S 2, we present a
height-integrated set of equations of the model. Self-similar
solutions are investigated in \S 3. The paper concludes with a
summary of the results in \S4.

\section{FORMULATION}
In order to implement thermal conductivity correctly it is
essential to know whether the mean free path is less than (or
comparable to) the scale length of the temperature gradient. For
electron mean free path which are greater than the scale length
of the temperature gradient the thermal conductivity is said to
'saturate' and the heat flux approaches a limiting value (Cowie
\& McKee 1977). But when the mean free paths are much less than
the temperature gradient the heat flux depends on the coefficient
of thermal conductivity and the temperature gradient. Generally,
thermal conduction transfers heat so as to oppose the temperature
gradient which causes the transfer. Menou (2005) discussed hot
accretion likely proceed under weakly-collisional conditions in
these systems. Thus, a saturated form of "microscopic" thermal
conduction is physically well-motivated, as we apply in this
study. However, one of the primary problems for studying the
effects of thermal conduction in plasmas is the unknown value of
the thermal conductivity.

We consider an accretion disk that is axisymmetric and
geometrically thin, i.e. $h/r<1$. Here $r$ and $h$ are,
respectively, the disk radius and the half-thickness. Our model
generalizes the usual slim disks around a neutron star or a black
hole (e.g., Muchotrzeb \& Paczy\'{n}ski 1982; Matsumoto et al.
1984; Abramowicz et al. 1988) by including the effect of thermal
conduction. The disk is supposed to be turbulent and possesses an
effective turbulent viscosity. We assume the generated energy due
to viscous dissipation and the heat conducted into the volume is
concerned are balanced by the advection cooling. Consider
stationary height-integrated equations described an accretion flow
onto a central object of mass $M_{\ast}$. In absence of mass
outflows, the continuity equation reads
\begin{equation}
\dot{M}=-2\pi r v_{\rm r} \Sigma,
\end{equation}
where $\dot{M}$ is the accretion rate, $v_{\rm r}$ is the
accretion velocity (and $v_{\rm r}<0$) and $\Sigma=2h\rho$ is the
surface density at a cylindrical radius. Also, $\rho$ is the
midplane density of the disk.

The equation of motion in the radial direction is
\begin{equation}
\Sigma v_{\rm r}\frac{d v_{\rm r}}{d
r}-\Sigma\frac{v_{\varphi}^2}{r}=-\frac{d P}{d
r}-\Sigma\frac{GM_{\ast}}{r^2},
\end{equation}
where $P=2h p$ is the integrated disk pressure written, in a form
compatible with radiation-dominated flow, as $P\simeq 2h a T^4/3$,
where $a$ is black body constant and $T$ denotes the midplane
temperature of the disk. Similarly, integration over $z$ of the
azimuthal equation of motion gives
\begin{equation}
r\Sigma v_{\rm r}\frac{d}{d r}(rv_{\varphi})=\frac{d}{d
r}[r^{3}\nu\Sigma\frac{d}{d r}(\frac{v_{\varphi}}{r})],
\end{equation}
where $\nu$ is a kinematic viscosity coefficient and we assume
\begin{equation}
\nu=\alpha c_{\rm s} h,
\end{equation}
where  $\alpha$ is a constant less than unity (Shakura \& Sunyaev
1973). Moreover, $h=(c_{\rm s}/v_{\rm K})r$, where $c_{\rm s}$ and
$v_{\rm K}$ are sound speed and the Keplerian velocity,
respectively: $c_{\rm s}=\sqrt{P/\Sigma}$ and $v_{\rm
K}=\sqrt{GM_{\ast}/r}$.

Now, we can write the energy equation considering the cooling and
the heating processes of the accretion flow. The locally released
energy due to viscous dissipation, $Q_{\rm vis}$, and the energy
transported by thermal conduction, $Q_{\rm cond}$, are balanced
by the advection cooling, $Q_{\rm adv}$. Thus,
\begin{equation}
Q_{\rm adv}=Q_{\rm cond}+Q_{\rm vis}.
\end{equation}
The advection cooling reads (Abramowicz et al. 1998)
\begin{equation}
Q_{\rm adv}=\frac{\dot{M}}{2\pi
r^{2}}\frac{p}{\rho}[\frac{4-3\beta}{\Gamma_{3}-1}\frac{d\ln T
}{d\ln r}+(4-3\beta)\frac{d\ln\rho}{d\ln r}],
\end{equation}
where $\beta$ is the ratio of the gas to the total pressure and
\begin{equation}
\Gamma_{3}=1+\frac{(4-3\beta)(\gamma -1)}{\beta+12(\gamma
-1)(\beta -1)}.
\end{equation}
Since we are considering the extreme case of radiation-dominated
accretion flow, we have $\beta=0$ and so $\Gamma_{3}=2/3$. Also,
we have

\begin{equation}
Q_{\rm vis}=\nu\Sigma r^{2}
[\frac{d}{dr}(\frac{v_{\varphi}}{r})]^{2},
\end{equation}
and (Cowie \& McKee 1977)
\begin{equation}
Q_{\rm cond}=-\frac{2h}{r^{2}}\frac{d}{dr}(r^{2}F_{\rm s})
\end{equation}
where as we have already mentioned $F_{\rm s}=5\Phi_{\rm s}\rho
c_{\rm s}^{3}$ is the saturated conduction flux on the direction
of the temperature gradient ($\Phi_{\rm s}<1$).
\section{SELF-SIMILAR SOLUTIONS}
The self-similar solution is not able to describe the global
behaviour of the accretion flow, because no boundary condition
has been taken into account. However, as long as we are not
interested in the the behaviour of the flow at the boundaries,
such solution describes correctly the true solution
asymptotically at large radii. We assume that each physical
quantity can be expressed as a power law of the radial distance,
 i.e.  $r^{\nu}$, where the power index $\nu$ is determined for
each  physical quantity self-consistently. The solutions are
\begin{equation}
\Sigma(r)=a \Sigma_{0}(\frac{r}{r_0})^{-1/2},
\end{equation}
\begin{equation}
v_{\varphi}(r)=b
\sqrt{\frac{GM_{\ast}}{r_0}}(\frac{r}{r_0})^{-1/2},
\end{equation}
\begin{equation}
v_{\rm r}(r)=- c
\sqrt{\frac{GM_{\ast}}{r_0}}(\frac{r}{r_0})^{-1/2},
\end{equation}
\begin{equation}
P(r)= d \frac{\Sigma_{0}GM_{\ast}}{r_0}(\frac{r}{r_0})^{-3/2},
\end{equation}
\begin{equation}
h(R)=f r_{0} (\frac{r}{r_0}),
\end{equation}
where $\Sigma_{\rm 0}$ and $r_{\rm 0}$ provide convenient units
with which the equations can be written in the non-dimensional
form. Thus, we obtain the following system of dimensionless
equations, to be solved for $a$, $b$, $c$, $d$ and $f$:
\begin{equation}
ac=\dot{m},
\end{equation}
\begin{equation}
-\frac{1}{2}ac^{2}-ab^{2}=\frac{3}{2}d-a,
\end{equation}
\begin{equation}
-\frac{1}{2}abc=-\frac{3\alpha}{4}\sqrt{\frac{d}{a}}fab,
\end{equation}
\begin{equation}
af^{2}-2d=0,
\end{equation}
\begin{equation}
\frac{3}{2}\dot{m}\frac{d}{a}=\frac{9}{4}\alpha\sqrt{\frac{d}{a}}fb^{2}a+5\Phi_{\rm
s}d\sqrt{\frac{d}{a}},
\end{equation}
where $\dot{m}=\dot{M}/(2\pi \Sigma_{0}\sqrt{GM_{\ast}r_{0}})$ is
nondimensional mass accretion rate. After some algebraic
manipulations we find
\begin{equation}
a=(\frac{2\sqrt{2}}{3\alpha}\dot{m})\frac{1}{f^{2}},\label{eq:a}
\end{equation}
\begin{equation}
b=\sqrt{1-\frac{3}{4}f^{2}-\frac{9}{16}\alpha^{2}f^{4}},\label{eq:b}
\end{equation}
\begin{equation}
c=(\frac{3\alpha}{2\sqrt{2}})f^{2},\label{eq:c}
\end{equation}
\begin{equation}
d=\frac{\sqrt{2}}{3\alpha}\dot{m},\label{eq:d}
\end{equation}
and $f$ is obtained from a forth order algebraic equation:
\begin{equation}
9 \alpha^{2} f^{4}+20 f^{2} - \frac{160\Phi_{\rm s}}{9\alpha} f -
16=0.\label{eq:main}
\end{equation}
We can solve this algebraic equation numerically and clearly only
the real root which corresponds to positive $b^{2}$ is a physical
solution.

If we neglect the thermal conduction (i.e., $\Phi_{\rm s }=0$),
the above equation reduces to a second order equation which can
be solved analytically (Wang $\&$ Zhou 1999; Shadmehri \&
Khajenabi 2005). In this case, the typical behaviour of the
solutions can be summarized as follows (Shadmehri \& Khajenabi
2005): (1) The surface density increases with the accretion rate,
and decreases with viscosity coefficient $\alpha$; (2) But the
radial velocity is directly proportional to $\alpha$; (3) The gas
rotates with sub-Keplerian angular velocity, more or less
independent of the coefficient $\alpha$; and (4) the opening
angle of the disk is fixed, independent of $\alpha$ and
$\dot{m}$. Note that Shadmehri $\&$ Khajenabi (2005)  applied a
diffusive prescription for the viscous stress tensor in their
analysis of the self-similar solution of optically thick
advection-dominated flows, but Wang $\&$ Zhou (1999) applied the
$\alpha p$ prescription. However, the typical behaviours of the
solutions are the same, irrespective of some differences in the
coefficients.

The equations (\ref{eq:a}), (\ref{eq:b}), (\ref{eq:c}),
(\ref{eq:d}) together with equation (\ref{eq:main}) describes
self-similar behaviour of the optically thick advection dominated
accretion disk with saturated thermal conduction. The algebraic
equation (\ref{eq:main}) shows that the variable $f$ which
determines the opening angle of the disk depends only on the
$\alpha$ and $\Phi_{\rm s}$. This behaviour is similar to the
case without thermal conduction. Thus, one can deduce that the
surface density and the pressure are directly proportional to the
mass accretion rate according to the equations (\ref{eq:a}) and
(\ref{eq:d}). But the rotational and the radial velocities are
both independent of the mass accretion rate (see equations
(\ref{eq:b}) and (\ref{eq:c})). Also, we can say that the
rotational velocity is sub-Keplerian according to the equation
(\ref{eq:b}).

\begin{figure}
\plotone{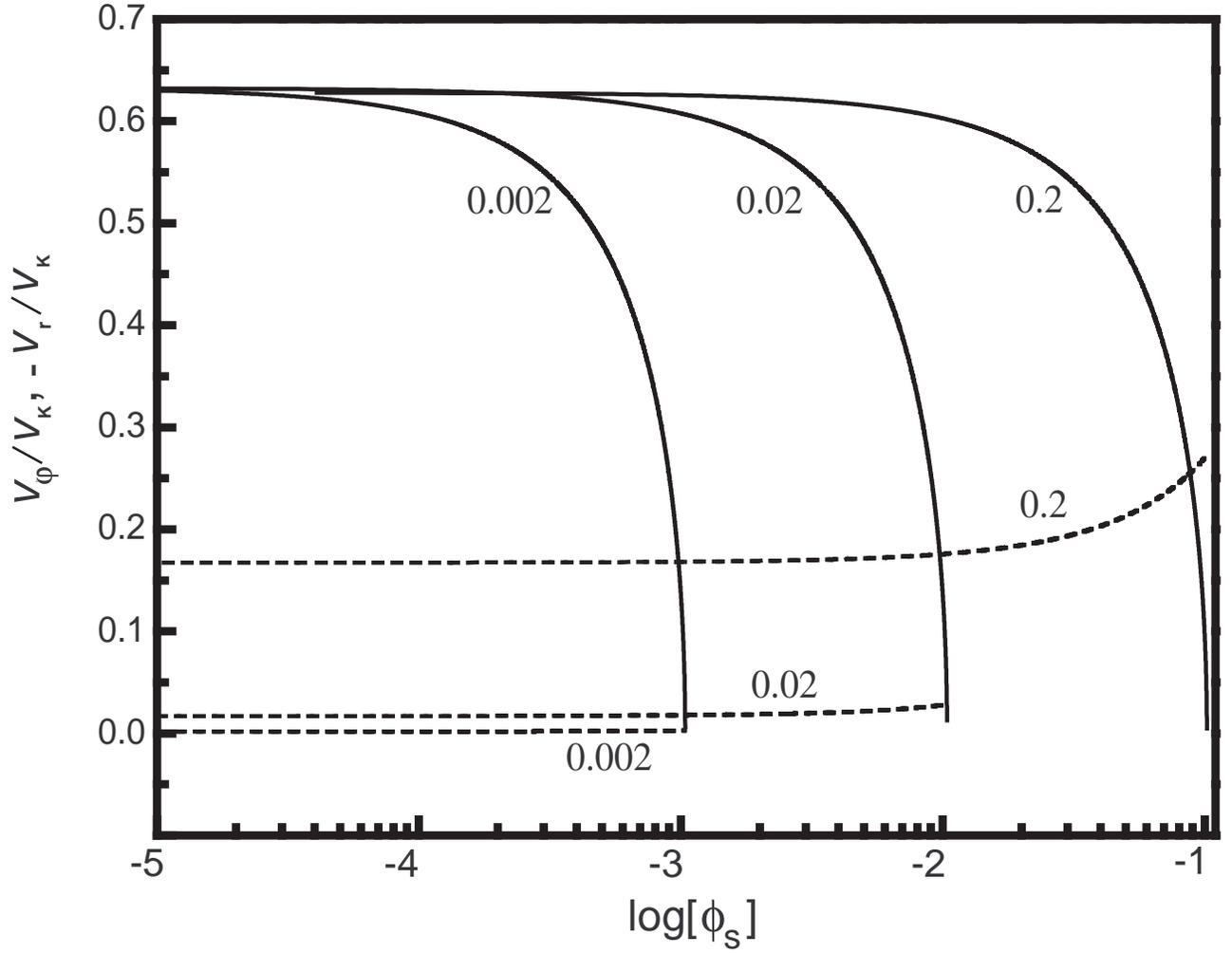} \caption{Profile of the  rotation velocity
($v_{\varphi}/v_{\rm K}$, solid lines) and the radial velocity
($v_{\rm r}/v_{\rm K}$, dashed lines) as a function of the
saturation constant, $\Phi_{\rm s}$. In order to make easier
comparison the rotational and the radial velocities are scaled by
the Keplerian velocity, $v_{\rm K}$. Each curve is labeled by
corresponding viscosity coefficient, $\alpha$.
}\label{fig:figure1}
\end{figure}

The dependence of the properties of the rotational and the radial
velocities on $\alpha$ and $\Phi_{\rm s}$ is illustrated in
Figure \ref{fig:figure1}. We consider three values of the
viscosity parameter, $\alpha=0.2, 0.02$ and $0.002$ for which the
velocities are plotted as a function of $\Phi_{\rm s}$. The solid
and dashed lines represent the ratios $v_{\varphi}/v_{\rm K}$ and
$v_{\rm r}/v_{\rm K}$, respectively. While the solution without
thermal conduction are recovered at small $\Phi_{\rm s}$ values,
we can see significant deviation as $\Phi_{\rm s}$ increases.
Clearly, the rotational velocity decreases with the magnitude of
conduction parameter. But for the radial velocity the behaviour
is different, i.e. the velocity increases as $\Phi_{\rm s}$
increases. For a fixed viscosity parameter $\alpha$, the solution
reaches the non-rotating limit at a specific value of $\Phi_{\rm s
}$ which we denotes by $\Phi_{\rm s}^{\rm c}$. If we extend our
solution to $\Phi_{\rm s}>\Phi_{\rm s}^{\rm c}$, the  equation
(\ref{eq:b}) gives negative $b^{2}$ which is clearly
unacceptable. The critical value $\Phi_{\rm s}^{\rm c}$ can be
calculated easily:
\begin{equation}
\Phi_{\rm s}^{\rm
c}=\frac{9}{10\sqrt{6}}\sqrt{\sqrt{4\alpha^{2}+1}-1}.
\end{equation}
The above equation shows the critical magnitude of conduction
parameter for which the solution tends to the non-rotating limit
depends only on the viscosity coefficient so that as $\alpha$
increases, the breakdown of the solutions occurs at larger values
of $\Phi_{\rm s}$. But as Menou (2005) mentioned in his analysis
of optically thin ADAFs with saturated conduction, we think this
behaviour is a simple pathological feature of 1D
height-integrated equations we solved.

Solutions of equation (\ref{eq:main}) show that for a fixed
$\alpha$, the variable $f$ increases with conduction parameter
$\Phi_{\rm s}$. Since the gas temperature varies directly in
proportion to $f$, we can say as the level of thermal conduction
is increased the temperature of flow in comparison to solutions
without thermal conduction increases. In fact, thermal conduction
behaves as an extra heating source. Also, equation (\ref{eq:a})
shows the surface density decreases as $\Phi_{\rm s}$ increases.
We can compare the relative values of advection cooling, viscous
heating and the transported energy by conduction using our
self-similar solutions. One can simply show that $Q_{\rm
vis}/Q_{\rm adv}=2(b/f)^{2}$ and $Q_{\rm cond}/Q_{\rm adv}=20
\Phi_{\rm s}/(9\alpha f)$. While both these ratios don't depend on
the mass accretion rate, the viscosity parameter $\alpha$ and the
thermal conduction parameter $\Phi_{\rm s}$ determine the
relative importance of the various physical processes in the
system. Viscous energy dissipation decreases as $\Phi_{\rm s}$
increases, but the contribution of the energy transported by the
conduction becomes more significant.

\section{Conclusions}

In this paper we have studied  an optically thick ADAFs with
thermal conduction. Considering the weakly-collisional nature of
hot accretion flows, a saturated form of thermal conduction has
been used as a possible mechanism of transporting of  energy.
Although our self-similar solutions have their own limitations,
they can illuminate some possible effects of the saturated thermal
conduction on radiation-dominated accretion flows. We can
summarize the main features of our solutions: (1) The surface
density is in proportion to the mass accretion rate and it
increases as the level of the thermal conduction is increased. (2)
The temperature of the gas increases with thermal conduction
parameter and the conduction term in the energy equation acts as
an extra source of heating. (3) The angular velocity is
sub-Keplerian, but it decreases with increasing thermal
conduction parameter so that it tends to a non-rotating limit.
(4) However, the radial velocity increases with thermal
conduction parameter. Considering these significant effects of
the thermal conduction,  global transonic solutions representing
optically thick (or thin) ADAFs with thermal conduction would be
an interesting topic for future studies and our results are a
step toward this goal.

\begin{references}

\reference{} Abramowicz, M. A., Chen, X., Kato, S., Lasota, J. P.,
Regev, O. 1995, ApJ, 438, L37

\reference{} Abramowicz, M. A., Czerny, B., Lasota, J. -P.,
Szuszkiewicz, E. 1988, ApJ, 332, 646

\reference{} Chen, X. 1995, MNRAS, 275, 641

\reference{} Cowie, L. L. $\&$ McKee, C. F. 1977, ApJ, 211, 135

\reference{} Honma, F. 1996, PASJ, 48, 77

\reference{} Igumenshchev, I. V., Abramowicz, M. A., $\&$
Narayan, R. 2000, ApJ, 537, L27

\reference{} Manmoto, T. et al. 2000, ApJ, 529, 127

\reference{} Matsumoto, R., Kato, S., Fukue, J., Okazaki, A. T.
1984, PASJ, 36, 71

\reference{} Menou, K. 2005, ApJL, submitted (astro-ph/0507189)

\reference{} Muchotrzeb, B. $\&$ Paczy\'{n}ski, B. 1982, Acta
Astron., 32, 1

\reference{} Narayan, R., Igumenshchev, I. V., $\&$ Abramowicz, M.
A. 2000, ApJ, 539, 798

\reference{} Narayan, R. $\&$ Yi, I. 1994, ApJ, 428, L13

\reference{} Shadmehri, M. $\&$ Khajenabi, F. 2005, MNRAS, 361,
719

\reference{} Shakura, N. I. $\&$ Sunyaev, R.A. 1973, A\&A,  24,
337

\reference{} Wang, J.-M. $\&$ Zhou, Y.-Y. 1999, ApJ, 516, 420






\end{references}
\end{document}